# Pedagogical approach to anomalous position and velocity

Younsik Kim[1], Suk Bum Chung[2,3] and Changyoung Kim[1,*]
[1]Department of Physics and Astronomy, Seoul National University, Seoul 08826, Korea
[2]Department of Physics and Natural Science Research Institute, University of Seoul, Seoul 02504, Korea
[3]School of Physics, Korea Institute for Advanced Study, Seoul, 02455, South Korea
* Corresponding Author: changyoung@snu.ac.kr

**Abstract:**
In this work, we discuss a pedagogical method in deriving the expressions for anomalous position and velocity. While we follow the steps used in optics in the derivation of the group velocity, we use Bloch wave functions instead of plane wave states. In comparison to the plane wave case, application of Bloch wave functions results in two additional terms in the expression of the group velocity: the Berry phase factor and anomalous position contributions. These two new terms with distinct origins eventually lead to the known anomalous velocity. Aiming for an intuitive understanding, we simulate the situation under an electric field using linear-combination-of-atomic-orbital states and visually demonstrate that the envelope function exhibits the transverse motion expected from an anomalous velocity.

## I.     Introduction

Electronic wave functions in solids deviate from those of free electrons due to the periodic potential from ions, and take the form of Bloch waves [1]. An important consequence of Bloch waves is that the deformation in a Bloch wave function in comparison to a free electron wave function, i.e., the periodic part of the wave function $u_{\bm{k}}(\bm{r})$, can result in (momentum space) Berry curvature. The Berry curvature acts as an effective magnetic field and thus induces a transverse motion, dubbed as anomalous velocity [2].

The Berry curvature and its associated anomalous Hall effect significantly influence both fundamental and applied scientific fields. In condensed matter physics, integrating the Berry curvature over the Brillouin zone yields the Chern number, a topological invariant that dictates key electronic and transport properties in two-dimensional systems [2]. This framework is essential for understanding phenomena such as the quantum Hall effect and various topological phases of matter [2]. On the applied side, Berry curvature is quintessential in orbitronics, particularly through the orbital Hall effect which involves the transverse flow of orbital angular momentum [3]. This effect is considered to be useful in the development of advanced electronic and spintronic devices, underscoring the impact of Berry curvature and anomalous velocity on both fundamental research and technological advances.

The concept and actual usage of the anomalous velocity have been around for many decades. Starting from the celebrated systems where the Hall conductivity is quantized [4,5], its derivations, partly due to overriding concern to avoid any accuracies, have been presented with high formal thoroughness [2,6,7], which regrettably have not offered an understanding as intuitive as the usual group velocity derivation in optics. It is easy to follow the formulae of the formal derivations note-by-note without obtaining the needed understanding of the physics. An intuitive understanding on the other hand often allows us to see hidden physics. Therefore, a more pedagogical approach to the problem from the wave function level may be desired.

The purpose of this work is to establish a pedagogical approach to the anomalous velocity (thus, the Berry curvature) at the level of introductory graduate quantum mechanics and solid-state physics. We will use an approach similar to what is commonly used in the demonstration of the group velocity in optics, but we will do it with Bloch wave functions instead of plane waves. The result shows that, upon application of Bloch wave functions, a couple of additional terms appear in the expression of the group velocity. These new terms with distinct origins constitute the anomalous velocity. Aiming for an intuitive understanding, we further present visual demonstration of the group velocity in two-dimension using linear-combination-of-atomic-orbital (LCAO) states.

## II. Derivation

### 1. Group velocity in optics

We first remind the readers how the group velocity of photons is obtained in optics. Consider the superposition of two traveling waves of equal amplitudes with a close proximity in $k$ and $\omega$.

$$e^{i((k+\delta k)x - (w+\delta w)t)} + e^{i((k-\delta)x - (w-\delta w)t)} = 2\cos(\delta k \cdot x - \delta w \cdot t) e^{i(kx-wt)} \quad (1)$$

The velocity of the cosine envelope function gives the group velocity which may be obtained by making the phase constant.

$$v_g = \frac{x}{t} = \frac{\delta w}{\delta k} \to \frac{\partial w}{\partial k} \quad \text{or} \quad \vec{v}_g = \vec{\nabla}_k \omega \quad (2)$$

While this works fine for light traveling in a uniform medium or free particles, its application to states in a solid-state system poses an immediate problem; an electronic state in a solid is not a simple plane wave but is a Bloch wave, which is a product of the plane wave and the function $u_k(r)$ that has the period of the crystal lattice. The approach of equation (1) is on a par with the semi-classical approach in solid state physics in which the property of $u_k(r)$ is mostly ignored. Therefore, we need to go through the same process but with Bloch wave functions.

### 2. Anomalous Position & Berry connection

Before we discuss the anomalous velocity, it is useful to touch upon the anomalous position and its connection to the Berry connection as it provides an important behavior without introducing the more complicated time dependence. We start with a time independent Bloch wave function,

$$\psi_{\vec{k}}(\vec{r}) = u_{\vec{k}}(\vec{r}) e^{i\vec{k}\cdot\vec{r}} \quad (3)$$

Let us consider two Bloch wave functions of equal amplitudes

$$\psi_{\vec{k}\pm\delta\vec{k}}(\vec{r}) = \left[ u_{\vec{k}}(\vec{r}) \pm \sum_\mu \delta k_\mu \frac{\partial u_{\vec{k}}(\vec{r})}{\partial k_\mu} \right] e^{i\vec{k}\cdot\vec{r}} e^{\pm i\delta\vec{k}\cdot\vec{r}} \quad (4)$$

We wish to investigate the sum of the two wave functions. Up to the 1$^{st}$ order in $\delta\vec{k}$, we have

$$\left| \frac{\psi_{sum}}{2} \right|^2 = \left| \frac{\psi_{\vec{k}+\delta\vec{k}} + \psi_{\vec{k}-\delta\vec{k}}}{2} \right|^2$$

$$= |u_{\vec{k}}|^2 \cos^2(\delta\vec{k}\cdot\vec{r}) + i \sum_\mu \delta k_\mu \left[ u_{\vec{k}}^* \frac{\partial u_{\vec{k}}}{\partial k_\mu} - u_{\vec{k}} \frac{\partial u_{\vec{k}}^*}{\partial k_\mu} \right] \cos(\delta\vec{k}\cdot\vec{r}) \sin(\delta\vec{k}\cdot\vec{r}) \quad (5)$$

We are interested in the variation of the envelope function in the length unit of the lattice constant. Therefore, we will integrate the equation over a unit cell, so that variation within a unit cell is ignored. Note that the trigonometric functions are slowly varying functions of $\vec{r}$ and do not change appreciably over a unit cell since $\delta\vec{k}$ is small. With

$$\langle |u_{\vec{k}}|^2 \rangle_{u.c.} = \langle u_{\vec{k}} | u_{\vec{k}} \rangle = 1$$

$$\langle i \sum_\mu \delta k_\mu \left[ u_{\vec{k}}^* \frac{\partial u_{\vec{k}}}{\partial k_\mu} - u_{\vec{k}} \frac{\partial u_{\vec{k}}^*}{\partial k_\mu} \right] \rangle = 2\delta\vec{k} \cdot i\langle u_{\vec{k}} | \vec{\nabla}_{\vec{k}} | u_{\vec{k}} \rangle = 2\delta\vec{k} \cdot \vec{\mathcal{A}}_{\vec{k}} \tag{6}$$

where the fact that the Berry connection $\vec{\mathcal{A}}_{\vec{k}}$ is always real has been used, equation (5) leads to

$$\langle \left|\frac{\psi_{sum}}{2}\right|^2 \rangle = \cos^2(\delta\vec{k} \cdot \vec{r}) + 2\delta\vec{k} \cdot \vec{\mathcal{A}}_{\vec{k}} \cos(\delta\vec{k} \cdot \vec{r}) \sin(\delta\vec{k} \cdot \vec{r})$$

$$= \frac{1}{2} + \frac{1}{2}\sqrt{1 + (2\delta\vec{k} \cdot \vec{\mathcal{A}}_{\vec{k}})^2} \cos(2\delta\vec{k} \cdot \vec{r} - 2\delta\vec{k} \cdot \vec{\mathcal{A}}_{\vec{k}}) \tag{7}$$

The position of the 'wave packet' may be obtained from a local maximum of the cosine function in equation (7) as is done in optics, that is,

$$2\delta\vec{k} \cdot \vec{r}_c = 2\delta\vec{k} \cdot \vec{\mathcal{A}}_{\vec{k}} \quad \text{or} \quad \vec{r}_c = \vec{\mathcal{A}}_{\vec{k}} \tag{8}$$

There are a couple of points to discuss. First of all, the result in equation (8) is the same as the anomalous position of a state $\vec{r}_{k,ano} = i\langle u_{\vec{k}} | \vec{\nabla}_{\vec{k}} | u_{\vec{k}} \rangle = \vec{\mathcal{A}}_{\vec{k}}$, discussed in the work on electric polarization by Vanderbilt [7,8]. The other is that it is generally orbital-dependent [9] and hence naturally related to the electric dipole moment described by $\vec{p} \approx -\alpha_K(\vec{k} \times \vec{L})$ discussed within the orbital angular momentum (OAM) picture [10,11,12]. Here, the shift in the electron density within a cell is equivalent to the shift in the envelope function of equation (7). It should be noted that $\vec{r}_{k,ano}$ should be proportional to $k$ for a momentum near a time reversal invariant momentum (TRIM) point [10].

### 3. Anomalous Velocity

For the anomalous velocity, we use time dependent wave functions. We recall that the time evolution of a Bloch wave function (with possibly time-dependent momentum) is given by [2,6]

$$\Psi_{\vec{k}}(\vec{r},t) = u_{\vec{k}}(\vec{r}) e^{i \int^t dt' (\vec{k}\cdot\dot{\vec{r}} - \widetilde{\omega}_{\vec{k}})} \tag{9}$$

where the frequency is effectively modified $\widetilde{\omega}_{\vec{k}} \equiv \omega_{\vec{k}} - i\dot{\vec{k}} \cdot \langle u_{\vec{k}} | \vec{\nabla}_{\vec{k}} | u_{\vec{k}} \rangle$, to account for the Berry phase factor $\exp(i\gamma) = \exp[\int d\vec{k} \cdot \langle u_{\vec{k}} | \vec{\nabla}_{\vec{k}} | u_{\vec{k}} \rangle] = \exp\left[\int^t dt' \dot{\vec{k}} \cdot \langle u_{\vec{k}} | \vec{\nabla}_{\vec{k}} | u_{\vec{k}} \rangle\right]$. Note that $\vec{k}$ depends on time and so does $\widetilde{\omega}_{\vec{k}}$.

We follow essentially the same steps discussed above but with time dependent wave functions.

$$\Psi_{\vec{k}\pm\delta\vec{k}}(\vec{r},t) = \left[u_{\vec{k}}(\vec{r}) \pm \sum_\mu \delta k_\mu \frac{\partial u_{\vec{k}}(\vec{r})}{\partial k_\mu}\right] e^{i \int^t dt' (\vec{k}\cdot\dot{\vec{r}} - \widetilde{\omega}_{\vec{k}})} e^{\pm i \int^t dt' (\delta\vec{k}\cdot\dot{\vec{r}} - \delta\widetilde{\omega}_{\vec{k}})} \tag{10}$$

Integrating $\left|\frac{\psi_{sum}}{2}\right|^2$ over a unit cell, we get

$$\langle \left|\tfrac{\Psi_{sum}}{2}\right|^2 \rangle = \cos^2\theta + 2\delta\vec{k}\cdot\vec{\mathcal{A}}_{\vec{k}}\cos\theta\sin\theta = \tfrac{1}{2} + \tfrac{1}{2}\sqrt{1+\left(2\delta\vec{k}\cdot\vec{\mathcal{A}}_{\vec{k}}\right)^2}\cos(2\theta-\phi) \qquad (11)$$

where $\theta = \int^t dt'\,(\delta\vec{k}\cdot\dot{\vec{r}} - \delta\widetilde{\omega}_{\vec{k}})$ and $\phi = 2\delta\vec{k}\cdot\vec{\mathcal{A}}_{\vec{k}}$. The first term in (11) gives the usual group velocity while the second term is new and does not exist in the semi-classical description of a particle. The second term provides the anomalous velocity.

Again, the position of the wave packet $\vec{r}_c$ is obtained from the local maximum of the cosine function in (11), that is, $2\theta - \phi = 0$.

$$2\int^t dt'\,(\delta\vec{k}\cdot\dot{\vec{r}} - \delta\widetilde{\omega}_{\vec{k}}) = 2\int^t dt'\,\tfrac{d}{dt'}\{\delta\vec{k}\cdot\vec{\mathcal{A}}_{\vec{k}}\}$$
$$\delta\vec{k}\cdot\dot{\vec{r}}_c - \delta\widetilde{\omega}_{\vec{k}} = \tfrac{d}{dt}\{\delta\vec{k}\cdot\vec{\mathcal{A}}_{\vec{k}}\} \qquad (12)$$

Then, the group velocity becomes what it should be [2,6],

$$\begin{aligned}
\dot{\vec{r}}_c &= \vec{\nabla}_{\vec{k}}\widetilde{\omega}_{\vec{k}} + \tfrac{d}{dt}\vec{\mathcal{A}}_{\vec{k}} = \vec{\nabla}_{\vec{k}}\left[\omega_{\vec{k}} - \dot{\vec{k}}_c\cdot\vec{\mathcal{A}}_{\vec{k}}\right] + \tfrac{d}{dt}\vec{\mathcal{A}}_{\vec{k}} \\
&= \vec{\nabla}_{\vec{k}}\omega_{\vec{k}} - \vec{\nabla}_{\vec{k}}\left(\dot{\vec{k}}_c\cdot\vec{\mathcal{A}}_{\vec{k}}\right) + \left(\dot{\vec{k}}_c\cdot\vec{\nabla}_{\vec{k}}\right)\vec{\mathcal{A}}_{\vec{k}} \\
&= \vec{\nabla}_{\vec{k}}\omega_{\vec{k}} - \dot{\vec{k}}_c\times\left[\vec{\nabla}_{\vec{k}}\times\vec{\mathcal{A}}_{\vec{k}}\right] = \vec{\nabla}_{\vec{k}}\omega_{\vec{k}} - \dot{\vec{k}}_c\times\vec{\Omega}_{\vec{k}}
\end{aligned} \qquad (13)$$

where a vector identity $\vec{A}\times(\vec{\nabla}\times\vec{B}) = \vec{\nabla}(\vec{A}\cdot\vec{B}) - (\vec{A}\cdot\vec{\nabla})\vec{B}$ was used. Note how the gauge dependences of $\vec{\mathcal{A}}_{\vec{k}}$ and $\widetilde{\omega}_{\vec{k}}$, both arising from the phase choice for $u_k(r)$, cancel each other out in the result for the group velocity.

### 4. Discussion

We note that there are two contributions to the anomalous velocity in equation (13). The first contribution is from $-\vec{\nabla}_{\vec{k}}\left(\dot{\vec{k}}_c\cdot\vec{\mathcal{A}}_{\vec{k}}\right)$ which originates from the Berry phase factor $\exp(i\gamma)$ in equation (9). As $\vec{k}$ changes over time, a wave function picks up a phase factor $\exp(i\gamma)$. The *difference* in the phase factor among the constituent waves of the wave packet determines the interference. Therefore, it is the momentum derivative of the phase factor that determines the position of the wave packet, which explains the presence of $\vec{\nabla}_{\vec{k}}$ in the term. This is a global effect from the overall phase of the wave function.

The other contribution $\tfrac{d}{dt}\vec{\mathcal{A}}_{\vec{k}}$, which is due to the anomalous position $\vec{r}_{k,ano} = \vec{\mathcal{A}}_{\vec{k}}$, is probably less familiar. As mentioned above, $\vec{r}_{k,ano}$ is dependent on $\vec{k}$. Therefore, the time evolution of $\vec{r}_{k,ano}$ arises from the time evolution of $\vec{k}$, which translates to an anomalous velocity. Since the *change* in $\vec{r}_{k,ano}$ directly translates to the anomalous velocity, we need $\tfrac{d}{dt}$ in the term. This is a local effect from displacement of an atomic wave function. It is worthwhile noting that the effect can be deduced from the original OAM work in which local displacement of an atomic wave function was approximated by $\vec{p} \approx -\alpha_K(\vec{k}\times\vec{L})$ [10,11,12]. It is also consistent with the result of a recent study on the anomalous position contribution to the orbital Hall effect [9].

## III. Visual demonstration

Microscopic visualization of the anomalous velocity in terms of two contributions described above has not been attempted so far, to the best of our knowledge. For a visual demonstration of the anomalous velocity, we try to use an LCAO state with a single atomic orbital at each lattice point,

$$\psi_{\vec{k}}(\vec{r}) = \frac{1}{\sqrt{N}} \sum_m e^{i\vec{k}\cdot\vec{R}_m} \phi_0(\vec{r} - \vec{R}_m) \qquad (14)$$

To make the task simple, we take the following considerations for the LCAO state. First, we may use atomic orbitals with finite OAM, e.g., hydrogen $p_x \pm ip_y$, for states with finite Berry curvature [10,13] along with the zero OAM $s$-orbital for zero Berry curvature states (see Fig. 1). As for the lattice, we use a 2-dimensional (2D) square lattice for simplicity. Finally, the direction of $\delta\vec{k}$ should be taken along the direction of the interest. That is, we take $\delta\vec{k}$ normal to $\vec{k}$ to investigate the group velocity component perpendicular to $\vec{k}$ (anomalous direction) by making the envelope function vary along the direction. On the other hand, $\delta\vec{k}$ should be taken parallel to $\vec{k}$ when the usual group velocity along $\vec{k}$ is investigated.

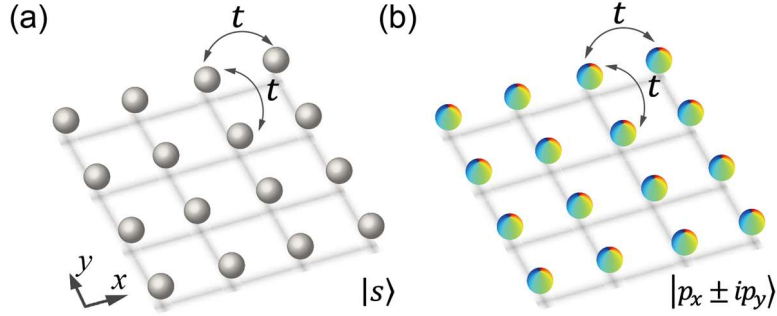

Fig. 1: Schematics for the 2D square lattice in the simulation. (a) $s$- and (b) $p_x \pm ip_y$ orbitals are used for zero and non-zero Berry curvature states, respectively. The phase of the $p_x \pm ip_y$ orbital is color coded, indicating non-zero orbital angular momentum of the orbital.

For the simulation, we utilized the PyTorch framework to perform parallelized calculations on a graphics processing unit [14]. Hydrogen atoms with either $1s$ (OAM = 0, Fig. 1a) or $2p_x \pm ip_y$ (OAM = $\pm\hbar$, Fig. 1b) orbitals are placed in a two-dimensional 20 × 20 square lattice on a 400 × 400 grid. The lattice constant is set to be $4a_0$ and $6a_0$ for OAM=0 and $\pm\hbar$, respectively, where $a_0$ is the Bohr radius. $|\delta\vec{k}|$ is taken to be $\frac{\pi}{20}$. For the time-dependence simulation, we set $k_x(t) = \frac{\pi}{8}(1 + \frac{t}{20})$.

### 1. Anomalous Position

We first demonstrate the anomalous position. Figure 2 depicts the electron density as well as the phase in the real space for various OAM and $\vec{k}$ values ($\vec{k}$ is along the positive $x$-direction). With zero OAM (no Berry connection), the electron density in Fig. 2a shows the local density maxima positioned at the atomic sites. As we change the OAM to $\hbar$ ($-\hbar$), the electron density shifts to the positive (negative) $y$-direction as shown in Fig. 2b (2c). Finally, we increase the $k$ value from $\pi/8$ to $\pi/4$ for $L = \hbar$ and the density further shifts to the positive $y$-direction as seen in Fig. 2d. This increased shift supports the notion that the anomalous position increases with $k$ and should be roughly proportional to $\Delta k$ referenced to a nearby TRIM point as mentioned above. These results are essentially reproduction of the previously reported work in Park *et al.* [10]. The shift of the electron density from the atomic site should represent the anomalous position.

Plotted in Figs. 2e-2h is the phase of the wave function. For the $L=0$ ($s$-orbital) case in Fig. 2e, the phase changes along the $x$-direction due to a finite $\vec{k}$ value. On the other hand, it not only varies along the $x$-direction but also rotates around each atom for the cases in Figs. 2f-2h due to the finite OAM value. The phase changes from the linear and angular momenta also interfere. Such an effect is especially visible for a large $\vec{k}$ value, seen as a less atomic-like phase profile in Fig. 2h.

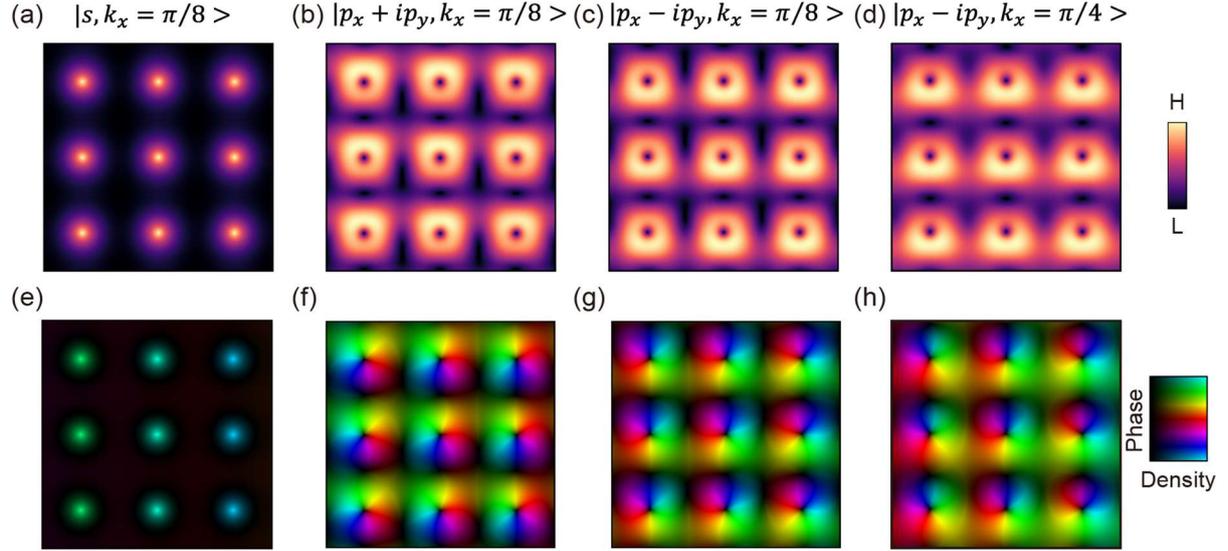

Fig. 2: Anomalous position. Electron density of the wave function for (a) $L=0$, $k_x=\pi/8$, (b) $L=+\hbar$, $k_x=\pi/8$, (c) $L=-\hbar$, $k_x=\pi/8$, (d) $L=+\hbar$, $k_x=\pi/4$,. The color scale is shown on the right-hand side of panel (d). (e)-(h) Color coded phase of the same wave functions. The color scheme is shown on the right.

2. **Anomalous velocity**

For a demonstration of the anomalous velocity, a time dependent wave function $\Psi_{\vec{k}}(\vec{r},t)$ should to be used. Using the LCAO wave function in equation (14), we obtain $\Psi_{sum} = \Psi_{\vec{k}+\delta\vec{k}} + \Psi_{\vec{k}-\delta\vec{k}}$ as described in equations (9) and (10). A few points need to be noted. Firstly, we use a 2D square lattice tight-binding model [12,13] for which the dispersion is given by $\omega_{\vec{k}} = -\alpha + 2t(\cos k_x a + \cos k_y a)$ where the hopping energy $t$ is appropriately adjusted in the simulation to have a discernable effect. In addition, the anomalous connection $\vec{\mathcal{A}}_{\vec{k}}$, calculated from the wave function of equation (14), is approximated to be $\vec{\mathcal{A}}_{\vec{k}} \approx \beta \vec{L} \times \vec{k}$ [13,15] in equations (9) and (10) for simplicity. Here, $\beta$ is a proportionality constant and $\vec{L}$ is the OAM of the atomic orbital (constant here). Finally, an electric field applied along the $x$-direction accelerates the electron and the momentum increases with time as described above ($\dot{k}_x = -eE$).

Plotted in Fig. 3 are time dependent real space electron density maps for various $L$ values. We first consider the $L=0$ (thus no Berry curvature) case. For the usual group velocity along the longitudinal direction, we set $\delta\vec{k}$ along the $x$-direction as shown in Fig. 3a. One can notice a few aspects from the time dependent maps. First of all, a constant phase (color coded) position changes with time, indicating a finite phase velocity. In this example, the phase velocity is positive but the phase moves to fast to observe a motion with the given time scale. Judging from the periodicity of the phase, the phase velocity increases with time as expected in the presence of an electric field. Moreover, looking at a node (dark line marked by the red arrow), the envelope function moves in the $x$-direction, showing that the longitudinal group velocity is positive. Even though not very obvious, a close look at the electron density shows that the electron density remains maximum at the atomic site.

Next, we set $\delta \vec{k}$ along the $y$-direction to investigate the transverse group velocity, for which the results are depicted in Fig. 3b. Note that we use a different time scale as the transverse group velocity is expected to be much smaller than the usual longitudinal group velocity. At $t = 0$, the envelope function has the maximum at $y = 0$ and nodes are located near the edge of the map (see the red arrow). At finite $t$ values, the phase again flows along the $x$-direction. However, the envelope function remains put with no apparent node motion, indicating that the transverse (anomalous) group velocity is zero without Berry curvature.

As we turn on the Berry curvature by changing $L$ to $\hbar$, the behaviors are different. In Fig. 3c, we plot the case for $L = \hbar$ with $\delta \vec{k}$ along the $y$-direction. The $t = 0$ map shows phase change along the $x$-direction as expected. In addition, the envelope function is also maximum at $y \approx 0$, similar to the case in Fig. 3b. As $t$ increases, a node of the envelope function marked by the red arrow moves in from the bottom of the image and continues to move upward with time, indicating a finite anomalous velocity along the $y$-direction. Finally, when we change the sign of the Berry curvature by flipping the OAM direction from $L = \hbar$ to $L = -\hbar$ as depicted in Fig. 3d, the anomalous motion also changes the direction to the negative $y$-direction. These observations visually demonstrate the behavior of the anomalous velocity from the intrinsic Berry curvature [16].

This work visually demonstrates the anomalous velocity. We hope that our work provides intuitive understanding of the anomalous velocity to researchers, especially to those who are new to the field. It is also worth mentioning that, even though this work serves the original goal, a better demonstration could be to show a transverse motion of a localized wave packet. Such work should require more complicated calculations with a need for more computing power, which is beyond what we are able to do at present. Therefore, we leave this for future efforts.

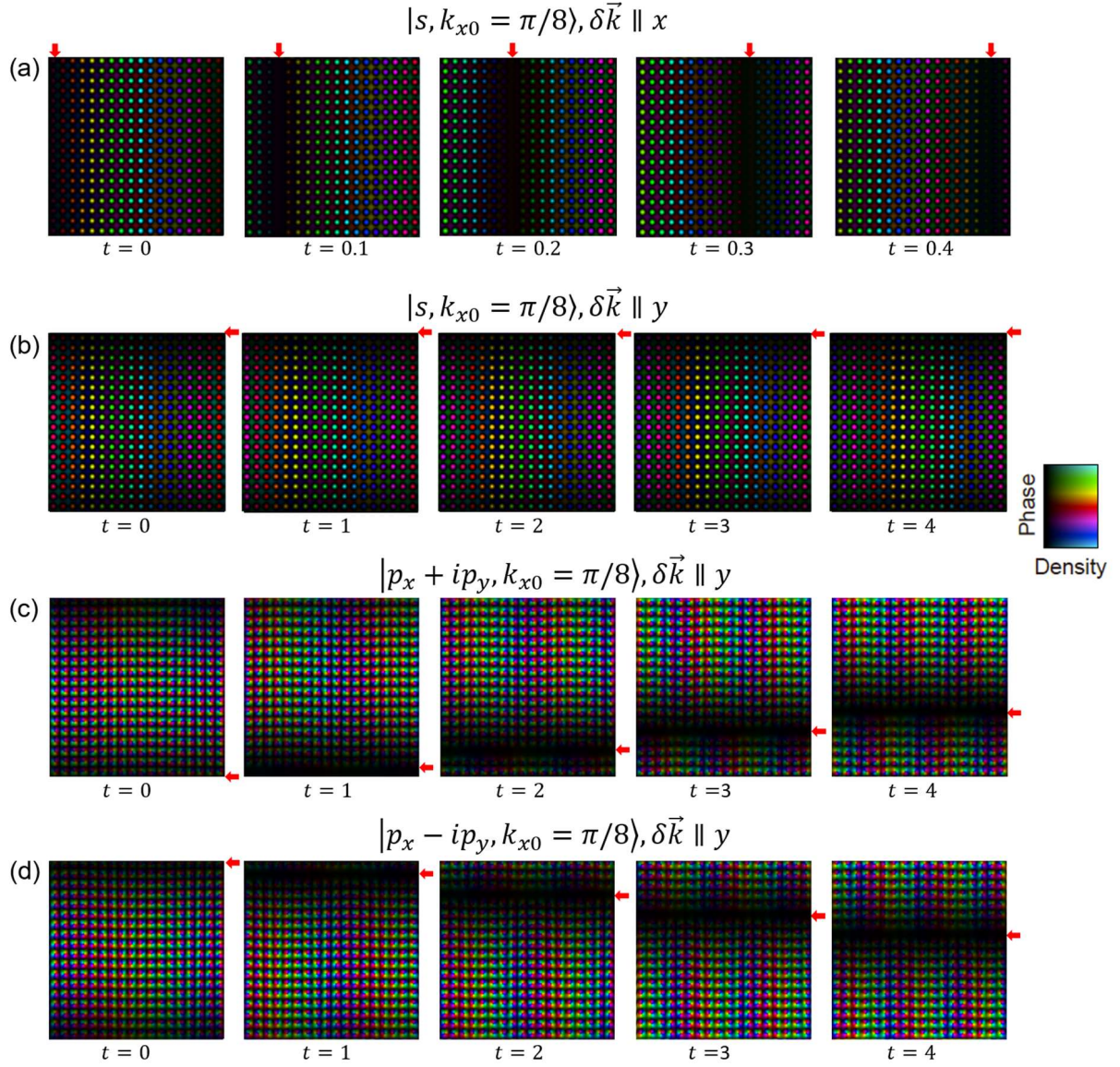

Fig. 3: (a) Time dependent density and phase map of the wave function for $L = 0$ with $\delta\vec{k}$ along the $x$-direction. Time dependent maps for (b) $L = 0$, (c) $L = \hbar$ and (d) $L = -\hbar$ with $\delta\vec{k}$ along the $y$-direction. $k_{x0}$ is fixed to $\pi/8$ throughout the figure. Red arrows indicate a node or a position related to a wave packet. Its motion shows the group velocity.

## Acknowledgements

This work was supported by the National Research Foundation of Korea (NRF) grant funded by the Korean government (MSIT) (No. 2022R1A3B1077234) and Global Research Development Center Cooperative Hub Program (GRDC) through NRF (RS-2023-00258359). SBC was supported by the National Research Foundation of Korea (NRF) grants funded by the Korea government (MSIT) (NRF-2023R1A2C1006144 and NRF-2018R1A6A1A06024977).